\documentstyle[12pt,a4wide]{article}
\def\l{\hat{L}_{n}}
\def\ld{\hat{L}^{\dagger}_{n}}
\def\qp{|{d\psi}\rangle}
\def\p{|{\psi}\rangle}
\def\h{{i\over\hbar}}
\def\bral{\langle \hat{L}_{n}\rangle_{\psi}}
\def\brald{\langle \hat{L}^{\dagger}_{n} \rangle_{\psi}}
\def\x{d\xi_{n}}
\def\numb{\overline{n}}
\begin{document}
\begin{flushright}
Imperial/TP/95--96/15
\end{flushright}
\vskip 1cm
\begin{center}
{\large{\bf Phase Space Localization and Approach to Thermal\\ Equilibrium
for a Class of Open Systems  }}
\vskip 2cm
{A.Zoupas\footnote{email: a.zoupas@ic.ac.uk. }
\vskip 0.4cm {\it Theoretical Physics Group,\\
Blackett Laboratory\\
Imperial College of Science, Technology \& Medicine\\
South Kensington, London  SW7 2BZ, U.K.}}
\vskip 0.7cm
{December 4, 1995}
\end{center}
\vskip 2cm


\begin{abstract}
We analyse the evolution of a quantum oscillator in a finite temperature
environment using the quantum state diffusion (QSD) picture. Following a
treatment
similar to that of reference \cite{HA.ZO}  we identify stationary solutions of
the corresponding It\^o equation. We prove their global stability and compute
typical time scales characterizing the localization process. The recovery of
the
density matrix in approximately diagonal form enables us to verify the approach
to thermal equilibrium in the long time limit and we comment on the connection
between QSD and the decoherent histories approach.
\end{abstract}
\section{Introduction}
\par
One of the approaches to quantum theory, developed the last years, is
the quantum state diffusion (QSD) picture
\cite{GiPe1,GiPe2,GiPe3}. The construction of this picture was
motivated by stochastic reduction theories and the necessity of
describing individual experimental outcomes. Mathematically it is
equivalent to the Lindblad master equation (hence it applies only to the
Markovian regime of open quantum systems). Its main feature is that it
describes the stochastic evolution of an individual system in Hilbert space,
a treatment complementary to the deterministic evolution provided by
the density matrix. Being a phenomenological picture, it has proved within its
domain of applicability to be in very good agreement with experiments involving
individual systems \cite{GKPT} and it is helping us to obtain more physical
intuition about processes of the microworld.

\par
Whenever the master equation for the reduced dencity matrix $\hat{\rho}$
describing
the evolution of the open system is of Lindblad form \cite{LIN} there is a
unique and
consistent way of unravelling it into an ensemble of individual system
state vectors each of which obeys a stochastic differential equation. Thus if
the evolution of $\hat{\rho}$ is given by,
\begin{equation}
 {d\hat{\rho} \over dt}=
{-}\h[\hat{H},\hat{\rho}]+\sum_{n}(\l\hat{\rho}\ld-{1\over 2}
 \ld\l\hat{\rho}-{1\over 2}\hat{\rho}\ld\l)
\end{equation}
the evolution for the state vector $\p$ of the individual system is  given by
the non-linear stochastic It\^o differential equation,
\begin{equation}
\qp=\h \hat{H}\p dt+\sum_{n}(\brald \l-{1\over 2}\ld\l-{1\over 2}
\brald \bral \p dt+\sum_{n}(\l -\bral )\p \x
\end{equation}
Here $\hat{H}$ is the Hamiltonian of the system in the absence of environment,
sometimes modified by terms depending on the Lindblad operators $\l$, $\ld$
which model the effect of the environment. The independent stochastic
differentials
$\x$ are complex and satisfy,
\begin{equation}
M\x =0,\;\;\; M(\x d\xi_{m})=0,\;\;\; M(\x d\xi^{\ast}_{m})=\delta_{mn}dt
\end{equation}
In the above $\langle  \rangle_{\psi}$ denotes mean value with respect to
$\p$ and M expresses the average over the ensemble. The density matrix is
recovered  by the formula
\begin{equation}
 \hat{\rho}=M(\p \langle \psi |)
\end{equation}
and consistency between the two pictures means that when  mean values
taken with respect to $\p$ in QSD are averaged over the ensemble, results
agree with those obtained using $\hat{\rho}$.
\section{The Model}
\par
In this paper we apply the QSD picture to the class of models
consisting of an harmonic oscillator in a finite temperature dissipative
environment. The Hamiltonian of the system in the absence of the environment is
given by,
\begin{equation}
H=\hbar \omega (\hat{\alpha}^{\dagger}\hat{\alpha}+{1\over{2}})
\end{equation}
while time evolution is described by a Lindblad master equation (1) with two
environment operators,
\begin{equation}
\hat{L}_{1}=[(\numb +1)\gamma ]^{1/2} \hat{\alpha},\;\;\;\;\;\;
\hat{L}_{2}=(\numb
\gamma )^{1/2}\hat{\alpha}^{\dagger}
\end{equation}
Here $\hat{\alpha}$ and $\hat{\alpha}^{\dagger}$ are the annihilation and
creation
operators for the harmonic oscillator,
\begin{equation}
\hat{\alpha} = (\sigma_{p}/ \hbar) \hat{q}+i(\sigma_{q}/ \hbar) \hat{p}
\end{equation}
where $\sigma_{q}$ and $\sigma_{p}$ denote the dispersions of position
and momentum for a standard coherent state,
\begin{equation}
\sigma_{q}=(\frac{\hbar}{2m\omega})^{1/2},\;\;\;\;\;\;\sigma_{p}=(\frac{\hbar m
\omega}{2})^{1/2}
\end{equation}
and
\begin{equation}
\numb =[\exp(\hbar \omega /kT)-1]^{-1}
\end{equation}
while the coefficient $\gamma$ represents dissipation.
The process under study could be the damping of a mode of the  electromagnetic
field in a cavity, coupled to a beam of two-level atoms, some of which are
initially excited or the damping of a coherent light beam propagating in a
weakly absorbing medium \cite{LA.PE}.
\par
The treatment will be similar to that of reference \cite{HA.ZO}. There, the QSD
picture was applied to quantum Brownian motion (QBM) models resulting
in approximate stationary solutions to It\^o equation. These were correlated
coherent
states. They minimize a more general uncertainty relation \cite{DOKU}
and are localized in phase space. For quadratic potentials the solutions
are exact and stability is global {\it i.e.} every initial state evolving
under It\^o takes the
form of the stationary one after some time. One of the main results of this
work was
that, after localization time, every initial density matrix tends to the form,
\begin{equation}
\hat{\rho} =\int f(p,q,t)|\psi_{pq}\rangle \langle \psi_{pq}|
\end{equation}
The above expression gives a natural way for approximately diagonalizing
$\hat{\rho}$ in the same basis  at all moments of time, expressing phase space
localization in the density matrix language. It resembles  the well
known P-representation of the density matrix \cite{KLASU} but differs in that
$|\psi_{pq}\rangle$ are correlated rather than standard coherent states and
$f(p,q,t)$ is always positive by construction. The expected approach to
thermalequilibrium was also proved in Ref. \cite{HA.ZO} and finally
localization rates were  studied and
the connection with decoherent histories was exemplified using phase space
projectors constructed from $|\psi_{pq}\rangle$.
\par
The linearity of Lindblad operators in position $\hat{q}$ and momentum
$\hat{p}$ makes our analysis resemble the one of QBM model but the physics
is obviously not the same. Here we are studying a process, Markovian at any
finite temperature T (even small) and  not only in the Fokker-Plank
limit (which is a high temperature limit), studied in
ref. \cite{HA.ZO}. The effect of the environment here is incorporated
into two Lindblad operators expressing different processes  and the
sources of phase space localization together with the localization
rates turn out to be of a different nature in both cases.
\par
The same system has also been subject of study previously \cite{GI,GISA,SGP}.
In \cite{SGP} the aproach to thermal equilibrium was shown numerically,
 while in \cite{GI} it was argued that we have coherent states as
stationary solutions to It\^o equation and a localization theorem was
proved in \cite{GISA}. Our study  concentrates on
diffferent aspects of the problem (and we also believe that the authors
of \cite{GISA} have made a mistake in their calculation). The purpose
of this paper is to study connections between localization and
decoherence as in ref. \cite{HA.ZO} but in a model different
to that one, and also to give an analytic account of the numerical
results of ref. \cite{SGP}.

\section{Stationary Solutions}
\par
Following [7] in the search for the stationary solution
we require that they satisfy the condition,
 \begin{equation}
\p +\qp = \exp[\h \hat{q} d\overline{p} -\h \hat{p} d\overline{q} +\h d\phi ]
\p
\end{equation}
with,
\begin{equation}
d\phi =\phi_{t} dt+\sum_{n=1}^{2} (\phi_{n} \x +\phi^{*}_{n} \x^{*})
\end{equation}
\begin{equation}
d\overline{p}= -mw^{2}\;\overline{q} dt-(\gamma/2)\overline{p}+
\sum_{n=1}^{2}[\sigma(\hat{p},\l)\x +\sigma(\l,\hat{p})\x^{*}]
\end{equation}
\begin{equation}
d\overline{q}= \frac{\overline{p}}{m} dt-(\gamma/2)\overline{q}+
\sum_{n=1}^{2}[\sigma(\hat{q},\l)\x +\sigma(\l,\hat{q})\x^{*}]
\end{equation}
Condition (10) means that the shape of the stationary solution is preserved
and only $\overline{p}$ and $\overline{q}$ change. Here $\phi$ is a
phase and by $\overline{p}$ and $\overline{q}$ we mean $\langle \hat{p}
\rangle_{\psi}$ and $\langle \hat{q}\rangle_{\psi}$ respectivelly.
We have also introduced:
\begin{equation}
\sigma(\hat{\Gamma},\hat{O}) = \langle\hat{\Gamma}^{\dagger}\hat{O}\rangle-
\langle\hat{\Gamma}^{\dagger}\rangle \langle \hat{O}\rangle
\end{equation}
After lengthy but not difficult calculations condition (9) leads to,
\begin{equation}
\langle q|\overline{p}\;\; \overline{q}\rangle \equiv \psi(q) =\exp[-\Lambda
(q-\overline{q} )^{2}+\h \overline{p} q]
\end{equation}
 with $\Lambda$ real,
\begin{equation}
\Lambda = \frac{1}{4\sigma^{2}_{q}} = {m\omega \over{2\hbar}}
\end{equation}
which is a standard (harmonic oscillator) coherent state  characterized by
minimum uncertainty:
\begin{equation}
\sigma^{2}_{q} \sigma^{2}_{p} = \hbar^{2}/4
\end{equation}
This property  makes the stationary solutions look like phase space points to
classical eyes.
\section{Localization Rates}
\par
The existence of stationary solutions gives rise to the questions of
stability and localization rates.
To prove that every state tends to the form of the stationary one, we observe
that
the stationary solution is eigenstate of $\hat{\alpha}$. To show
global stability, we need to show that,
\begin{equation}
Md{(\Delta \hat{\alpha})}^{2} \leq 0
\end{equation}
Now for a non Hermitian operator $\hat{O}$ we can define the spread as,
\begin{equation}
(\Delta \hat{O})^{2} \equiv \sigma(\hat{O},\hat{O})
\end{equation}
Using the two above equations we obtain,
\begin{equation}
Md{(\Delta \hat{\alpha})}^{2}=\frac{\gamma}{2\hbar^{2}}(\numb +1/2)[\hbar^{2}
 -4R^{2}-2\frac{\sigma^{2}_{q}}{\sigma_{p}^{2}}(\Delta \hat{p})^{4}
-2\frac{\sigma^{2}_{p}}{\sigma_{q}^{2}}(\Delta \hat{q})^{4}]
\end{equation}
where R, is the symmetrized correlation between $\hat{q}$ and $\hat{p}$,
\begin{equation}
R\equiv \frac{1}{2}\langle \{\hat{p},\hat{q} \} \rangle -\langle \hat{p}
\rangle
\langle \hat{q} \rangle
\end{equation}
 while $(\Delta \hat{p})^{2}$ and
$(\Delta \hat{q})^{2}$ denote the dispersions. Then setting,
\begin{equation}
(\Delta \hat{q})^{2}= \sigma^{2}_{q}(1+Q)\;\;\;\;\;\;\;\;\;\;\;\;\;\;
(\Delta \hat{p})^{2} = \sigma^{2}_{p}(1+P)
\end{equation}
clearly
\begin{equation}
P\geq -1,\;\;\;\;\;\;\;\;\;\;\;\; Q\geq -1
\end{equation}
and stationary solution is obtained for $P=Q=R=0$. The dispersion of
$\hat{\alpha}$ reads in terms of P and Q,
\begin{equation}
(\Delta \hat{\alpha})^{2} = (P+Q)/4 \geq 0
\end{equation}
which substituted in (20) leads to
\begin{equation}
Md{(\Delta \hat{\alpha})}^{2}=-2\gamma (\numb+1/2)[\frac{R^{2}}{\hbar^{2}}
+\frac{P^{2}}{8}+\frac{Q^{2}}{8}+ (\Delta \hat{\alpha})^{2}]
\end{equation}
The above expression is negative and vanishes only for a coherent
state. This proves the global stability of the stationary
solutions. Our expression for the rate of localization is not in agreement with
the one obtained in \cite{GISA}.
\par
It is obvious from (25) that the localization
process is characterized by a timescale of order $t_{l}=[\gamma
(\numb+1/2)]^{-1}$. This coresponds to a minimum rate of localization,
\begin{equation}
M\frac{d(\Delta \hat{\alpha})^{2}}{dt}\leq -2\gamma(\numb+1/2)(\Delta
\hat{\alpha})^{2}
\end{equation}
Then using expression (8) we obtain,
\begin{equation}
t_{l}\sim \frac{1}{\gamma}\tanh(\hbar \omega/2kT)
\end{equation}
which for $\hbar \omega << kT$ gives,
\begin{equation}
t_{l}\sim \frac{\hbar \omega}{\gamma kT}
\end{equation}
while for $\hbar \omega >> kT$ we have,
\begin{equation}
t_{l}\sim \frac{1}{\gamma}
\end{equation}
When studying the approach to thermal equilibrium the meaning of
(30) will become evident.
In the case of an initial state consisting of a superposition of
wavepackets a large distance $d$ apart, $(\Delta \hat{q})^{2}\sim
d^{2}$ and term $Q^{2}$ will be the dominant one in (25). Then in the
high temperature limit, we obtain by virtue of (22) and (24),
\begin{equation}
t_{l}\sim \frac{\hbar^{2}}{d^{2}m \gamma kT}
\end{equation}
This result agrees with the usual ``decoherence time'' for high $T$ \cite{PAZU}
\section{Thermal Equilibrium}
\par
The approach to equilibrium in Ref. \cite{SGP} was shown by computing
numericaly the behaviour of $\langle n \rangle$ and the occupancy
probabilities for various number states $|n \rangle$. Using our stationary
solution we may do the same analitically.For times greater than the
localization time, $\langle\hat{n(t)}\rangle_{\psi}$ is given by,
\begin{equation}
\langle\hat{n(t)}\rangle_{\psi} \equiv \langle\hat{n(t)}\rangle_{st}=
\frac{\sigma^2_{p}}{ \hbar^2}\; \overline{q}^2+\frac{\sigma^2_{q}}{
\hbar^2}\; \overline{p}^2 = \alpha^{*}(t) \alpha(t)
\end{equation}
Here $\alpha$ denotes the eigenvalue of $\hat{\alpha}$. Using
equations (13) and (14) we may derive an evolution equation for
$\alpha(t)$ \cite{GAR},
\begin{equation}
\alpha(t)= \alpha(0)\exp[-(i \omega + \gamma/2)t)+\sqrt{\numb \gamma}
\int_{0}^{t}\exp[-(i \omega + \gamma/2)(t-t')]d \xi_{2}(t')
\end{equation}
It is clear that in the long time limit only the part of the solution
with the integral over the stochastic process will survive.
Then $\langle \hat{n}(t) \rangle$ reads,
\begin{equation}
\langle \hat{n}(t)\rangle = \numb \gamma
\int_{0}^{t}\exp[(\frac{\gamma}{2}+i\omega)\tau
+(\frac{\gamma}{2}-i\omega)\tau']d\xi_{2}(\tau)d\xi_{2}^{*}(\tau').
\end{equation}
{}From the above expression only when we take the mean over the ensemble
we are able to say that once equilibrium is reached $\langle
\hat{n}\rangle$ will fluctuate around $\numb$ as expected. Then the
standard result,
\begin{equation}
M\langle\hat{n}(t)\rangle_{st}= \numb
\end{equation}
is recovered,since
$M[d\xi_{2}(\tau)d\xi_{2}^{*}(\tau')]=\delta(\tau-\tau')d\tau d\tau'$.
Using our stationary solution, in
principle, we are able to compute time averages over the stochastic
process and thus reproduce  the numerical result of Ref.
\cite{SGP} for the occupation probabilities analytically.
(This result is needed to exhibit the thermal nature of the
fluctuations in (34)).
However, the manipulations such a computation would involve are
cumbersome. We shall then proceed using the density operator to treat
this problem.
We may recover $\hat{\rho}$ following the treatment of Ref
\cite{HA.ZO}. Once localization has taken place $\rm Eq^n$ (4) is of
the form,
\begin{equation}
\hat{\rho} =\int f(p,q,t)|pq\rangle \langle pq|
\end{equation}
Hence every initial $\hat{\rho}$ approaches the phase space diagonal
form (36) on the localization timescale.
Equation (36) is the P-representation of $\hat{\rho}$ with the nice property
that
$f(p,q,t)$ is positive. A Fokker-Plank equation is easily derived for
$f(p,q,t)$ \cite{GAR}:
\begin{equation}
\frac{\partial}{\partial t}f =-{p\over{m}}\frac{\partial}{\partial q}f+
{\gamma \over{2}}\frac{\partial}{\partial q}(qf)+m\omega^{2} q\frac{\partial}
{\partial p}f+{\gamma \over{2}}\frac{\partial}{\partial p}(pf)+{\hbar
\overline{n}
 \gamma \over{2m\omega}}\frac{\partial^{2}}{\partial q^{2}}f+{\hbar \over{2}}
\overline{n} \gamma m\omega \frac{\partial^{2}}{\partial p^{2}}f
\end{equation}
It admits a unique Gaussian stationary solution being globally stable and
approached for times t $>>\gamma^{-1}$
\begin{equation}
f_{s}(p,q)= \frac{1}{2 \pi \numb} exp(-{1\over{\overline{n}}} | \alpha | ^{2})
\end{equation}
This is the
P-symbol expression for a thermal density operator. Hence every
initial state approaches thermal $\hat{\rho}$ for long times, as
expected. Then the elements $\langle n| \hat{\rho}|n \rangle$
may be evaluated and give the standard result,
\begin{equation}
\langle n| \hat{\rho}|n \rangle = \frac{1}{\numb}(1+ \numb^{-1})^{-(n+1)}
\end{equation}
for the occupation probabilities, since the probability of observing
$n$ quanta in a coherent state({\it i.e.} $|\langle n|pq\rangle |^2$)
is given by the poissonian distribution with mean $|a|^{2}$.
We note that our result together with that of Ref. \cite{SGP}
(which is numerical and for a specific initial state) give the first
traces of a proof that the ergodic hypothesis holds for our system.
\par
One of the featurs of our model is that it is applicable at low
temperatures as well.When the temperature
 $T\rightarrow 0$, we see from equations (8), (33) and (36) that the system
decays to its ground state as expected. (for zero temperature
 $f_{s}(p,q)$ becomes infinitely sharp around $\alpha= 0$ as
seen from (38). A very interesting result obtained  in connection with
equation (30) is that localization and
relaxation to thermal equilibrium proceed essentialy on the same
time scale for $T\rightarrow 0$. This should be expected since for
$\numb =0$, $\hat{L}_{1}=\gamma \hat{\alpha}$ and $\hat{L}_{2}=0$,
expressing only dissipation of energy to the environment.
Therefore in this limit the mechanism of decoherence
is not very efficient, exhibiting the deeper connection
between environmentaly induced decoherence and noise.

\section{Comments}
\par
In the context of explaining the emergence of classical behaviour as a result
of the interaction of a system with its environment, the approaches
of ``decoherence of density operators'' \cite{ZEH,HU,PAZU,ZU1} and of
``decoherent histories'' \cite{GE.HA,GRIF,OMNES,DOWK,JJ} give the
frameworks for a systematic treatment. In the former the approximate
diagonilization of the reduced density operator, expressed by (36) for this
class of models is essential while in the latter the decoherence of histories
is the
prerequisite to emergent classicality. The close connection between
QSD and decoherent histories \cite{DGHP} can be illustrated using the
results obtained so far. The localization properties of  the
stationary solutions to
It\^o equation makes natural the study of phase space histories. We
thus take phase space projectors of the form,
\begin{equation}
\hat{P}_{pq}=\int_{{\Gamma}_{pq}}dpdq\;\;|pq\rangle \langle pq|
\end{equation}
Here $\Gamma_{pq}$ is a phase space cell. Obviously those projectors are only
approximate ones, the validity of approximation depending on the area covered
by the cell and the nature of its boundary \cite{OMNES}.
Construct now histories with the above projectors and assume our initial state
is $\hat{\rho}$, then the decoherence functional is given by,
\begin{equation}
D(\underline{a}, \underline{a'}) = Tr(P_{a_{n}}K^{t_{n}}_{t_{n-1}}
[P_{a_{n-1}}....K^{t_{2}}_{t_{1}}[P_{a_{1}}K^{t_{1}}_{t_{0}}[\rho_{0}]
P_{a'_{1}}]....P_{a'_{n-1}}]P_{a'_{n}})
\end{equation}
Here $\underline{a},\underline{a'}$ denote the two strings of
projections at succesive moments of time $t_{1}....t_{n}$, $a_{k}$
characterizes the phase space cell at a time moment $t_{k}$  and
$K^{t_{k}}_{t_{k-1}}$
is the propagator of the density operator obeying the master equation
of our model. $D(\underline{a}, \underline{a'})$ given by (42) is
recovered  from the decoherence functional of the closed system after
having traced out the environment and under the assumptions that the
initial density operator factorizes and  the process is Markovian.
\par
For  intervals between projections longer than
the localization time the density operator will always evolve to the
diagonal form (37). Studying then (42) we conclude that because of the
diagonality of $\hat{\rho}$  the off- diagonal elements
$D(\underline{a}, \underline{a'})$ will be approximatelly equal to zero.
Therefore approximate decoherence is achieved. Then probabilities
computed are found to be most strongly peaked around the classical path
and are essentialy the same for both approaches provided we always
refer to scales grater than $\frac{\hbar}{2}$. This is because the
probabilities of  phase space histories (computed from the diagonal
elements of $D$) and the probabilities in QSD are both equivalent
to the probabilities derived from the description of our system
using the Fokker-Planck equation (38).
Thus all the
conclusions of the analysis of Ref. \cite{HA.ZO} are valid here as well.
However, a comment we would like to make is that in our model localization and
decoherence occur in the Lindblad operator while in Ref. \cite{HA.ZO}
do not. This is due to the fact that here the eigenstates of
$\hat{L}$ are preserved by the Hamiltonian while in general this need
not be the case.
\par
In connection with the results obtained in \cite{HA.ZO} we can then conclude
that the diagonality of $\hat{\rho}$ in a basis consisting of eigenstates
of the same operator at all momements of time, the construction of
decohering phase space histories (requirments for emergent classicality)
and the study of the approach to thermal equilibrium
may be acomplished in a natural way using the stationary solutions to
It\^o equation.
Hence the description of processes in terms of individual systems, the
localization properties of its solutions
and the close connection with decoherent histories,
make quantum state diffusion very useful both on intuitive
and calculational grounds in the study of the emergence of classical
behaviour of open quantum systems.

{\bf Acnowledgements}\\
I would like to thank J.J.\ Halliwell for pointing out this problem to
me and for useful suggestions. I would also like to thank  B.\ Garraway,
C.\ Anastopoulos, and Todd Brun for stimulating discussions.

\end{document}